\documentclass[aps,prb,twocolumn]{revtex4-1}
\usepackage{amssymb}
\usepackage{graphicx}
\usepackage{hyperref}

\begin{document}
	
\title{Fermi surface topology and large magnetoresistance in the topological semimetal candidate PrBi}

	\author{Amit Vashist $^{1,*}$, R. K. Gopal$^{1,*}$, Divya Srivastava$^{2,3}$, M. Karppinen$^2$, and Yogesh Singh $^1$}
	\affiliation{$^1$ Department of Physical Sciences,
		Indian Institute of Science Education and Research Mohali,
		Sector 81, S. A. S. Nagar, Manauli, PO: 140306, India.}
	
		\affiliation{$^2$ Department of Chemistry and Materials Science , Aalto University, FI-00076, Aalto, Finland}
\affiliation{$^3$ Department of Physics, Central University of Rajasthan,  Ajmer 305 817, Rajasthan, India}
	\date{\today}

\begin{abstract}
We report a detailed magnetotransport study on single crystals of PrBi.  The presence of $f$-electrons in this material raises the prospect of realizing a strongly correlated version of topological semimetals.  PrBi shows a magnetic field induced metal insulator transition below $T \sim 20$~K and a very large magnetoresistance ($\approx 4.4 \times 10^4~\%$) at low temperatures ($T= 2$~K). We have also probed the Fermi surface topology by de Haas van Alphen (dHvA) and Shubnikov de Haas (SdH) quantum oscillation measurements complimented with density functional theory (DFT) calculations of the band structure and the Fermi surface.  Angle dependence of the SdH oscillations have been carried out to probe the possible signature of surface Dirac fermions. We find three frequencies corresponding to one electron ($\alpha$) and two hole ($\beta$ and $\gamma$) pockets in experiments, consistent with DFT calculations. The angular dependence of these frequencies is not consistent with a two dimensional Fermi surface suggesting that the transport is dominated by bulk bands.  Although the transport properties of this material originate from the bulk bands, the high mobility and small effective mass are comparable to other compounds in this series proposed as topologically nontrivial. 

	\end{abstract}
	\maketitle	
	
\section{Introduction}
Topological materials, such as topological insulators\cite{1,2}, Weyl\cite{3,4} and Dirac semimetals\cite{5,6} present the opportunity of studying relativistic Dirac like low energy excitations in condensed matter systems.  Due to their unique bulk band topology, such topological semi-metals show exotic physical properties such as extremely high magnetoresistance (XMR), a very high mobility along with low carrier density, chiral anomaly driven negative MR and a Berry phase of $\pi$.  A topic of immense current interest is trying to combine topological character with strong electronic correlations.  The Kondo insulator SmB$_6$ \cite{20} and the pyrochlore iridates $R_2$Ir$_2$O$_7$ \cite{Wan2011} are examples of materials which are candidates to show exotic properties from a combination of Topology and electron correlations. Common among these materials is the presence of $4f$-electrons.  The rare-earth monopnictides $RX (R =$ rare-earth, $X = $ Sb, Bi) are candidate materials for hosting correlated electrons with Topological properties.  Indeed unusual magneto-transport properties similar to those observed in TSMs have been reported for some rare earth based monopnictides  \cite{7,8,9,10,11,12}.  However, the topological character of the bands participating in transport is still a matter of debate with contrasting reports about this issue. For instance, the electronic band structure of LaBi is regarded as topologically nontrivial, consisting of two Dirac cones at the X point, whereas other groups claim one Dirac cone at X point from ARPES studies.  Another report has shown that the bands near X point have a Dirac nodal line nature instead of Dirac cones\cite{13, 14}. Several other compounds in this series, such as LaAs, LaSb, PrSb, ScSb, YSb, TmSb, and LuBi \cite{7,11,15, 16, 17, 18,19} show similar transport properties as LaBi, despite the fact that all of them have been reported to be topologically trivial.  It would be interesting to study the $RX$ materials with a magnetic rare-earth since the $4f$-correlated electrons might influence the Topological properties.  Most the the magnetic $RX$ materials however show long-ranged magnetic order at low temperatures \cite{Tsuchida-a} hindering a study of low temperature magneto-transport properties.  PrBi is special in this context since it has a partially filled $4f$-shell but does not order at least down to $1.8$~K \cite{Tsuchida-b}.   
	
A recent ARPES study on PrBi has revealed a nontrivial Z2 band structure topology in the bulk characterized by an odd number of Dirac cones \cite{13,14,15}.  On the other hand, the iso-structural materials PrSb and ScSb are found to be topologically trivial by band structure calculations, partly owing to the reduced spin orbit coupling due to replacement of Bi by Sb in these materials.  PrBi is thus a novel material where the presence of $f$-electrons in addition to the possible topological band structure found in ARPES raises the prospect of realizing a strongly correlated version of topological semimetals.

In this work, we investigate in detail the low temperature magneto-transport properties of single crystals of PrBi, a rare earth monopnictide consisting of partially filled $f$-electron shells. Magnetic measurements are consistent with a trivalent Pr$^{3+}$ valence state and reveal paramagnetic temperature dependent behaviour down to a temperature of $T = 2$~K\@.  Systematic temperature dependent Hall measurements on these single crystals show nonlinear variation with magnetic field, suggesting a compensated nature of charge carriers in PrBi.  The temperature dependent resistivity in zero magnetic field shows a typical semimetallic character and a field induced metal insulator transition is observed at low temperatures $T \leq 20$~K\@.  We observe a large magnetoresistance ($\approx 4.4 \times 10^4~\%$) at low temperatures ($T= 2$~K) in a magnetic field $B = 9$~T\@.  Additionally, pronounced quantum oscillations are observed in the magnetoresistance and magnetization data.  We found two frequencies in Shubnikov de Haas (SdH) and three frequencies in de Haas van Alphen (dHvA) oscillations.  We map out the Fermi surface (FS) of PrBi by angle dependent quantum oscillation measurements which reveal a three dimensional FS.  
We use density functional theory (DFT) to calculate the band structure and Fermi surface of PrBi.  Although, the frequencies for quantum oscillations predicted by DFT are overestimated, we find qualitative agreement between DFT and experiments.  A Landau fan diagram is constructed from the experimental quantum oscillation data and this is used to estimate the Berry curvature $\phi$.  We estimate a non-trivial Berry curvature of $\pi$ for PrBi suggesting a connection between the unusual magneto-transport properties of PrBi and other mono-pnictides to a topological band structure.

\begin{figure*}[t]
\includegraphics[width=1\textwidth]{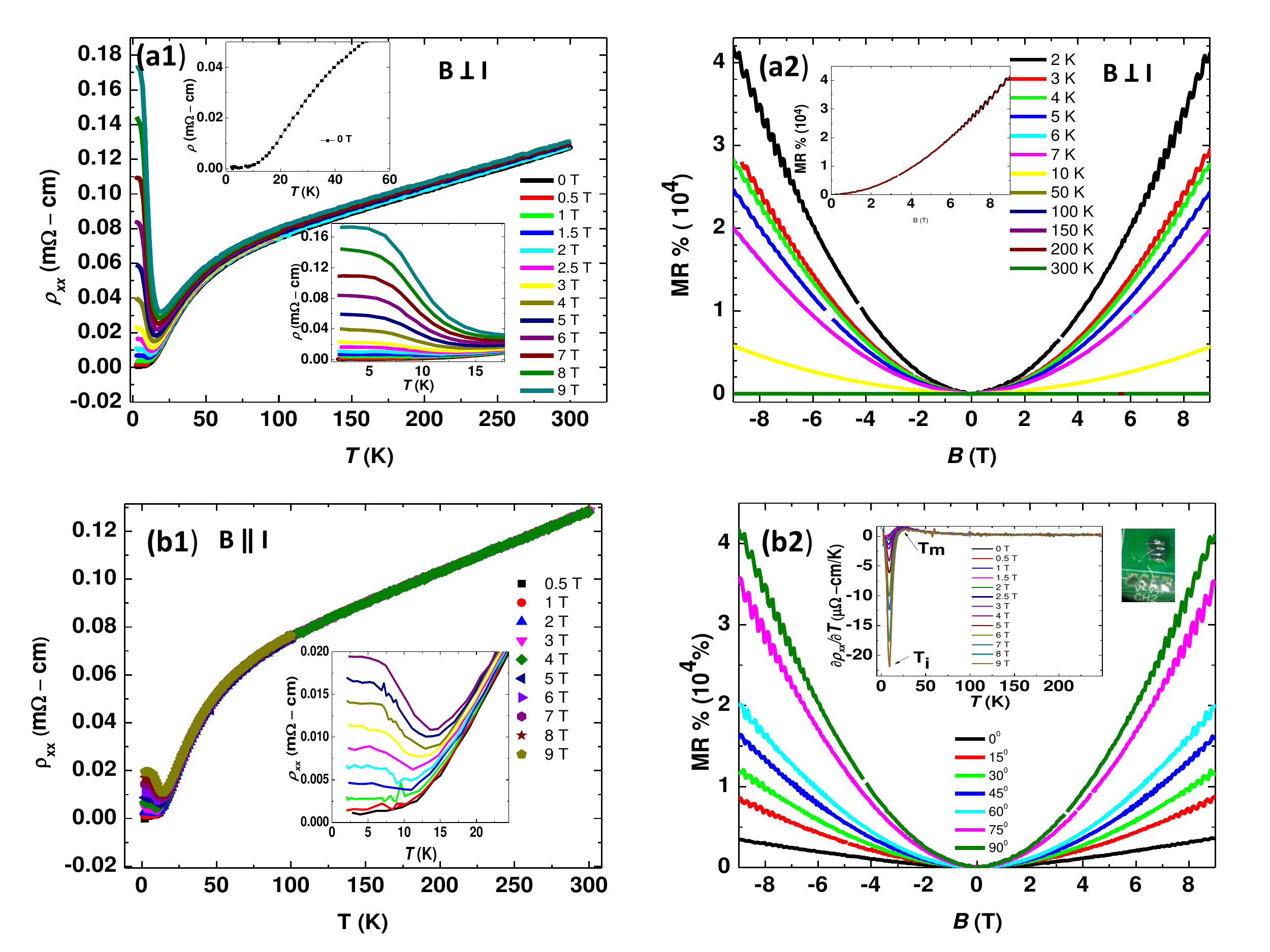}
\caption{(a1 and b1) Temperature dependence of $\rho_{xx}$ in the two configurations $B\perp I$ and $B\parallel I$, at various magnetic field. Upper inset of (a1) shows the zero field behaviour of $\rho_{xx}$ at low temperature.  Lower inset of (a1 and b1) show enlarged view of $\rho_{xx}$ at low temperature, at various fields. (a2) Magnetic field dependence of $\rho_{xx}$ in the direction $B\perp I$ at various temperatures.  Inset shows the power law behaviour of MR.  (b2) Magnetic field dependence of $\rho_{xx}$ at $2$~K by varying the (out of plane) angle between $B$ and $I$.  Inset shows $d\rho_{xx}/dT$ versus temperature at various fields. Top right shows an image of the crystal with electrical contacts for magneto-transport measurements.} 
\label{Fig-RES}
\end{figure*}
	
\section{Experimental Details}
High quality single crystals of PrBi were synthesized using indium flux. The starting elements Pr (99.9 \%), Bi (99.999 \%), and In (99.999 \%) were taken in molar ratio of $1:1:20$. The elements were placed into an alumina crucible and sealed in an evacuated quartz tube. The sealed quartz tube was heated to a temperature of $1050~^o$C in $10$~h, kept at this temperature for $5$~h  and then slowly cooled to $700~^o$C at a rate of $2~^o$C/hr. At this temperature excess In was removed by a centrifuge. The resulting crystals were of typical dimensions 3.5 $\times$ 2 $\times$ 1 mm. The structure of the crystals was confirmed by powder x-ray diffraction on crushed crystals.  The chemical composition of the crystals was verified by energy dispersive spectroscopy (EDS) using a scanning electron microscope. The electrical resistance, angle dependent magnetoresistance and magnetic susceptibility measurements were performed using a Quantum Design physical property measurement system (QD-PPMS) in the temperature range $2$ to $300$~K and in magnetic fields of up to $9$~T\@.

\section{Computational Details}
The density functional theory calculations in the GGA framework were carried out using plane-wave and pseudo potential methods as implemented in the Quantum Espresso package \cite{Giannozzi2009, Giannozzi2017}. A kinetic-energy cutoff of 65 Ry was used for the 
plane-wave expansion of valence wave-functions. Electronic structure was determined with and without spin-orbit coupling. The Perdew-Burrke-Ernzerhof (PBE) \cite{PBE} exchange-correlation functional was used with fully relativistic and scalar-relativistic ultra-soft pseudopotentials for spin-orbit interaction (SOI) and non-spin-orbit interaction (non-SOI), respectively. For the self-consistent calculations a $24\times 24 \times 24$ k-point mesh was used to sample the reduced Brillouin zone.
A finer k-pint mesh of $50 \times 50 \times 50$ was used to generate the Fermi surface (FS). The FS was visualized using the XCRYSDEN software\cite{Kokalj2003}. The maximally localized Wannier function method \cite{Pizzi2014} were employed to calculate the dHvA frequencies using the SKEAF code \cite{Rourke2012}. 

\section{Experimental Results}

\begin{figure*}[t]
\includegraphics[width=1\textwidth]{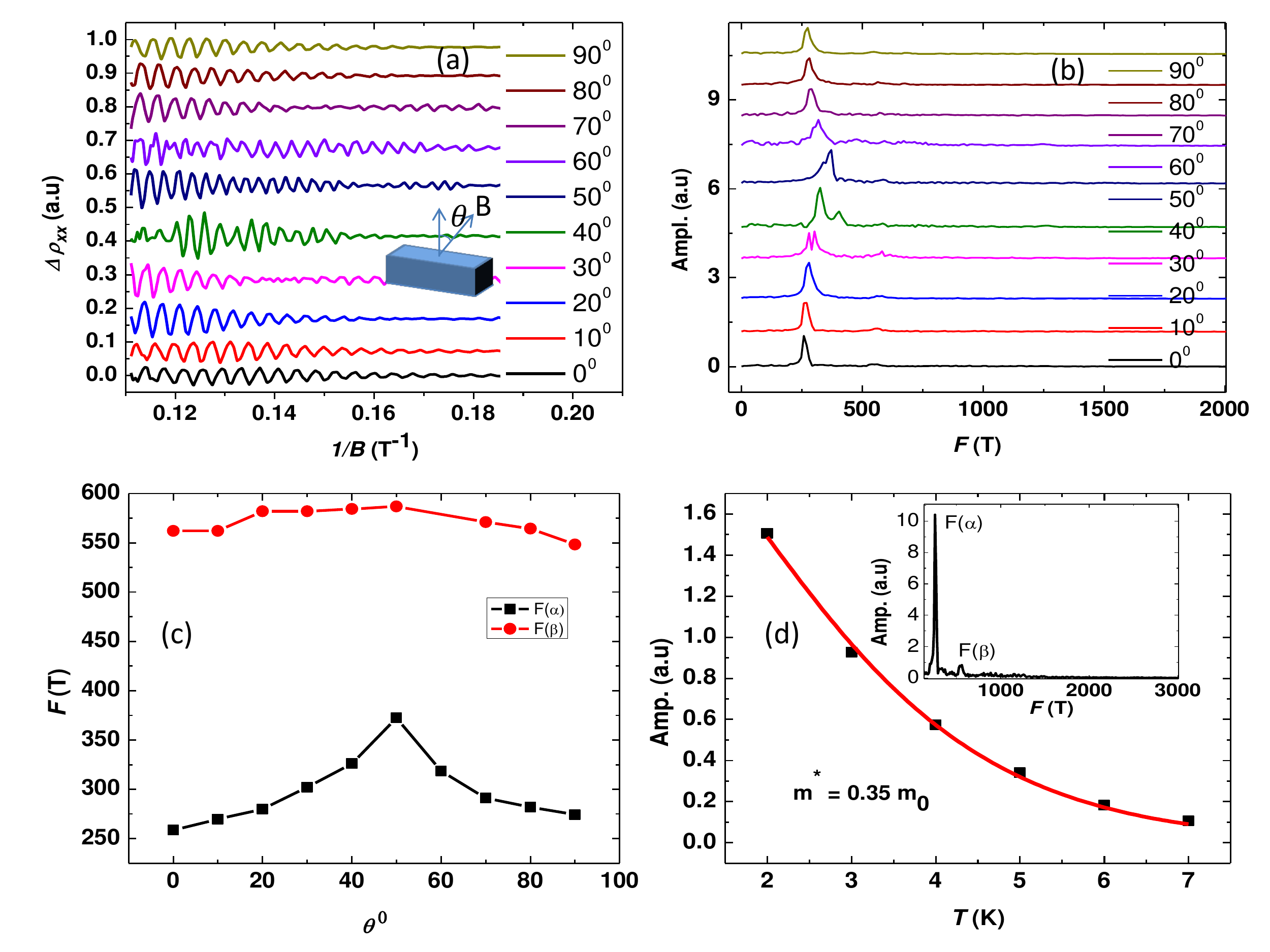}
\caption{(a) Background subtracted SdH oscillations at various angles, at 2 K. (b) The FFT spectra at different angles. (c) The angular dependence of the frequencies $F(\alpha)$ and $F(\beta)$. (d) Fitting of the temperature dependent amplitude in the FFT spectra by the L-K formula (see text for details).}
\label{Fig-SdH}
\end{figure*}	
	
The temperature ($T$) dependent resistivity ($\rho(T)$) and magnetoresistance (MR) data on the PrBi single crystals with the magnetic field parallel ($B \parallel I$) and perpendicular ($B \perp I$) to the current $I$, are shown in Fig.~\ref{Fig-RES}.  A large value of the zero field residual resistivity ratio $\rho(300$~K$)/\rho(2$~K) $= 374$, indicates the high quality of the single crystals used in this study. The zero field resistivity follows a linear $T$ dependence from $300$~K to $100$~K as can be seen in Fig.~\ref{Fig-RES}(a1). Below $T=100$~K, the resistivity falls rapidly and nearly saturates in the low temperature regime $T \leq 10$~K\@. 

The application of magnetic fields drives a metal to insulator like transition (MIT) at low temperatures for both perpendicular (B $\perp$ I) and in-plane magnetic field (B $\parallel$ I) orientations as can be seen from Figs.~\ref{Fig-RES}(a1) and (b1).  This MIT like feature in $\rho(T)$  is more prominent in the B $\perp$ I configuration and exceeds the room temperature value of resistivity in the low $T$ regime at a field of $9$~T (see Fig.~\ref{Fig-RES}(a1)).  The $\rho(T)$ in magnetic field goes through a minimum at a temperature ($T_m \sim 15$~K), shows a sharp upturn below $T_m$, before eventually saturating at lower temperatures.  This low temperature behaviour of $\rho(T)$ in fields is highlighted in the lower insets of Figs.~\ref{Fig-RES}(a1) and (b1).  The minimum in $\rho(T,H)$ can also be seen in the $d\rho(T)/dT$ vs $T$ plot shown in inset of Fig.~\ref{Fig-RES}~(b2).  Similar field driven MIT features have been observed in the topological semimetals WTe2, NbP, and TaAs \cite{23,24,25}. 

Since the $\rho(T)$ appears to have a larger response for $B \perp I$, we show the MR for the $B \perp I$ configuration at various temperatures in Fig.~\ref{Fig-RES}~(a2).  The MR for $B \parallel I$ is qualitatively similar but smaller in magnitude.  The MR for temperatures above $T_m$ is quite small.  For lower temperatures, the MR increases dramatically.  The MR at $2$~K in a field of $9$~T reaches a very large value of $\sim$ $4.2 \times 10^4~ \%$ which is comparable to values observed in other mono-pnictide compounds \cite{8,11,12}. The MR has a near parabolic $B$ dependence and is nonsaturating up to $9$~T\@. We successfully fit the MR with a power law dependence, MR $\propto B^m$ (inset of Fig.~\ref{Fig-RES}~(a2)) with an exponent $m = 1.82$.  This nearly quadratic behaviour is a characteristic feature of three dimensional charge transport in multiband metals.  Below $5$~K SdH quantum oscillations riding on top of the smooth quadratic MR can be clearly observed at higher fields and these oscillations gets stronger with decreasing temperature as can be seen in Fig.~\ref{Fig-RES}~(a2).  

The angular dependence of the MR at $T = 2$~K is presented in Fig.~\ref{Fig-RES}(b2) as the out of plane angle between the magnetic field $B$ and the current $I$ is varied starting from $B || I$ for which the angle is $0^o$.  It can be seen that the MR for $B || I$ is an order of magnitude smaller than for $B \perp I$.  The MR for all angles shows a quadratic $B$ dependence.  Additionally, quantum oscillations can be seen in the MR data at all angles.  
	
\begin{figure*}[t]
\includegraphics[width=1\textwidth]{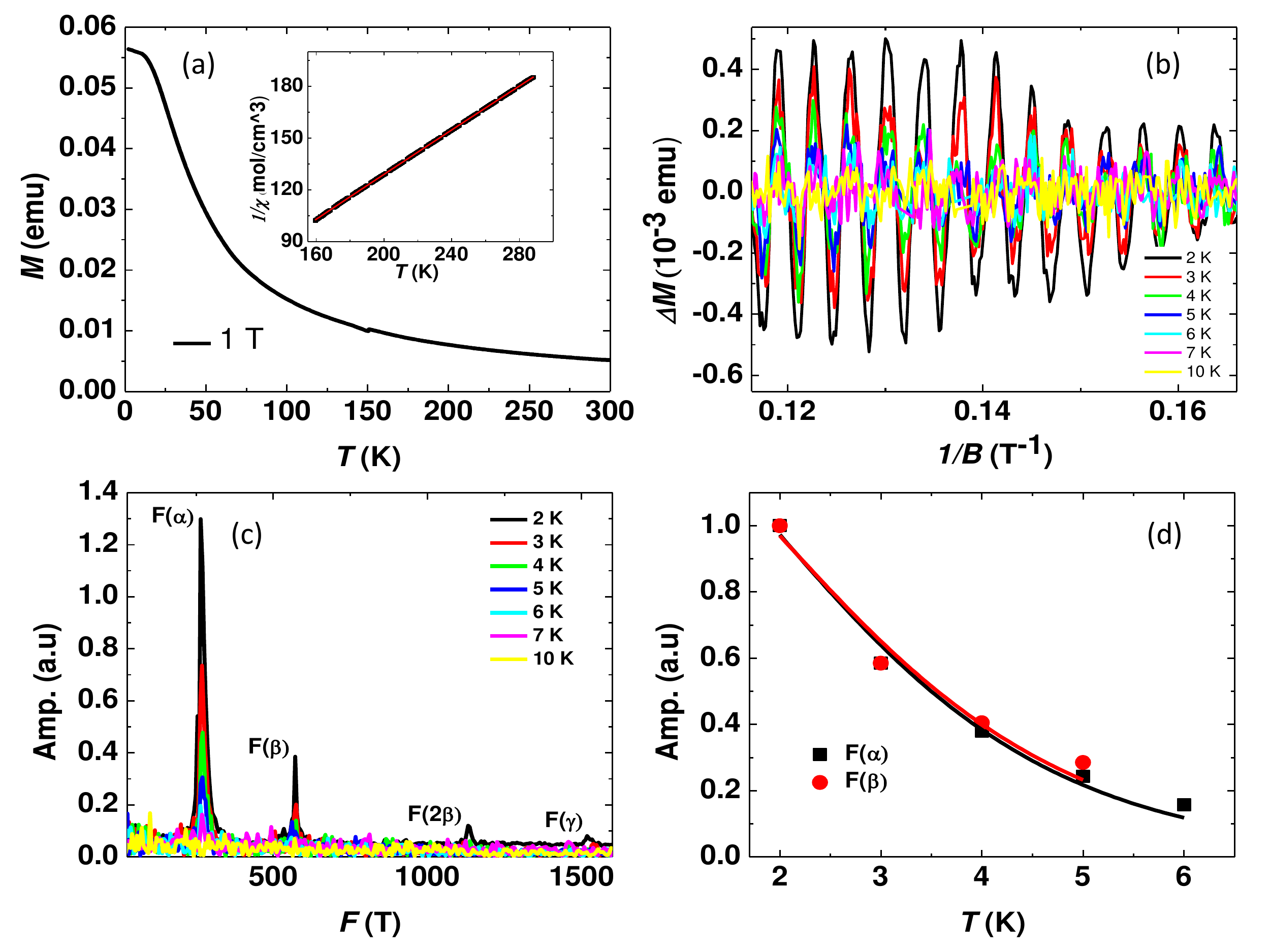}
\caption{(a) Temperature dependence of the magnetic moment of PrBi in an applied field of 1 T. (b) The temperature dependent dHvA oscillations obtained after background subtraction. (c) Temperature dependence of the FFT spectra. (d) Fitting of temperature dependent FFT spectra using L-K formula. } 
\label{Fig-MH}
\end{figure*}
	
We next study the Fermi surface topography of PrBi by measuring the angle dependence of the quantum oscillations in the MR data, taken at $2$~K\@. In order to extract clear oscillations, a polynomial background has been subtracted from the MR at each angle. The resulting periodic oscillations at several angles is plotted against inverse of the magnetic field B in Fig.~\ref{Fig-SdH}(a). It is clear from this data, that the oscillations do not vanish and remain pronounced as we rotate the magnetic field from the B $\perp$ I to the B $\parallel$ I configuration. This suggests that the charge carriers in PrBi are of three dimensional character.  A fast Fourier transform (FFT) of the quantum oscillation data in Fig.~\ref{Fig-SdH}(a) for each angle was done and is shown in Fig.~\ref{Fig-SdH}(b).  The FFT shown in Fig.~\ref{Fig-SdH}(b) for different angles are shifted by arbitrary amounts for clarity.  From the FFT of the data we find two fundamental frequencies, which for $B \perp I$ (angle $= 0^o$) are $F(\alpha) = 258.49$~T and $F(\beta) = 561.94$~T (inset  Fig.~\ref{Fig-SdH}(d)). From Fig.~\ref{Fig-SdH}(b) the angular evolution of the two frequencies $F(\alpha)$ and $F(\beta)$ can be seen.  With increasing angle we observe a gradual shifting of the $\alpha$ frequency towards higher values. In addition, a new frequency emerges as the tilt angle reaches $\theta = 30^{0}$. This frequency vanishes or merges into another frequency as we approach $\theta = 50^{0}$. This type of frequency shifting/splitting has been observed in other mono-pnictides like TmSb and CeSb suggesting a similar Fermi surface \cite{26, 27, 28}.  This angular variation has been proposed to result from the quasi two dimensional character of the Fermi surface in the case of LaBi \cite{9}. However, the observed variation in frequency can also result from an anisotropic ellipsoidal Fermi surface \cite{30, 31}, which is most likely the case here due to the $\alpha$ electron pockets at the $X$ point in the Brillouin zone. 
	
We have plotted the values of the two main frequencies as a function of the angle in Fig.~\ref{Fig-SdH}(c). From the angular variation, it can be seen that the frequency $F(\alpha)$ shows a monotonic increase in value up to an angle $\theta = 50^{0}$, and decreases thereafter reaching close to its $\theta = 0^o$ value at $\theta = 90^o$. This behaviour is consistent with an anisotropic elliptical Fermi surface orbit for the $\alpha$ bands. The other frequency $F(\beta)$ remains almost unchanged in the entire variation in the tilt angle. This frequency corresponds to the hole pockets centered at the $\gamma$ band in the first Brillouin zone. We thus conclude that the quantum oscillations observed in the PrBi single crystals primarily originate from the three dimensional bands instead of two dimensional ones. In addition to this, we have measured the MR in the longitudinal magnetic field configuration B $\parallel$ I to look for any signature of the chiral anomaly induced negative MR. We did not detect any negative MR.

\begin{table*}	
\caption{Fermi surface parameters for PrBi obtained from the SdH and dHvA data shown in Figs.~\ref{Fig-SdH}~ and ~\ref{Fig-MH} }.
			\scalebox{1}{
				\begin{tabular}{|c|c|c|c|c|c|c|c|} 
					\hline
					&	F (T) & $A_f (\AA^{-2})(10^{-2})$ & $K_f (\AA^{-1}) (10^{-2})$ & $m^*/m$ & $v_f (m/s) (10^{5})$ & $ E_f (meV)$ &$ n (cm^{-3}) (10^{19})$\\ [0.ex] 
					\hline
					$SdH $&258.49 & $2.474$ &  $8.876 $& 0.35 & $2.93$  & $170.21$ &2.36 \\ 
					\hline
					&561.94 & $5.378$  &$13.08$&$-$& $-$& $-$& 7.56\\
					\hline
					dHvA&	264.40 & $2.530$ &  $8.980 $& 0.32 & $3.24$ & $190.70$ &2.40\\ 
					\hline
					&571.19 & $5.466$  &$13.20$& 0.31 & $4.9$  & $423.90$ &7.70\\
					\hline
					&1517.9 & $14.52$  &$21.50$& $-$ & $-$  & $-$ &33.60\\
					[1ex] 	\hline
					
				\end{tabular}}
				\label{Table 1}
			\end{table*}

We have also carried out temperature and magnetic field dependent magnetization measurements on the same PrBi crystal. The temperature dependence of the magnetization measurement reveals the paramagnetic nature of PrBi (see Fig.~\ref{Fig-MH} (a)). The plateau below $10$~K is consistent with previous reports\cite{Tsuchida-b, 11,32} and most likely arises from Van Vleck temperature independent paramagnetism.  We did not observe any signature of long range ordering down to $2$~K and by fitting the data above $160$~K to a Curie-Weiss law (inset Fig.~\ref{Fig-MH} (a)) we estimated an effective magnetic moment $\mu_{eff} = 3.42~\mu_B$.  This value is close to the value expected for trivalent Pr and is consistent with a previous report \cite{Tsuchida-b}.  

The high quality of our PrBi single crystals allowed us to observe pronounced quantum oscillations (dHvA) in magnetization measurements as well and are shown in Fig.~\ref{Fig-MH} (b). The FFT of these oscillations reveals three fundamental frequencies.  The frequencies F$(\alpha) = 264.4$~T and F$(\beta) = 571.2$~T are close to the frequencies found from the SdH oscillation data presented above.  An additional frequency (apart from F$(2\beta) = 1131.8$~T, the second harmonic of the $\beta$ band) F$(\gamma) = 1517.9$~T is also observed in the $T = 2$~K data which has a very small amplitude which is most likely the reason it was not observed in the SdH oscillations.  For this reason we are unable to track the temperature dependence of the frequency of the $\gamma$ band.

We next calculate the transport parameters corresponding to the two main frequencies F$(\alpha)$ and F$(\beta)$ found in the SdH and dHvA oscillations. The area of the extremal orbits corresponding to these two frequencies, can be calculated using the Onsager relation, F$ = A_{F} (\varphi/2\pi^2 )$, where $\varphi = h/e$, is the flux quantum, $A_{F}$ is Fermi surface area, $h$ is the Plank constant and $e$ is the charge of an electron. By assuming a circular cross section of the cyclotron orbits at the Fermi level, we have calculated the values of the Fermi momentum $k_{f}$ and Fermi energy $E_{f}$.  The calculated area of the Fermi surface for the two frequencies extracted from SdH and dHvA oscillations are listed in Table~\ref{Table 1}. The area/carrier concentration corresponding to the hole orbit ($\beta$ pockets) is much larger than the electron orbit ($\alpha$ pockets), which implies the dominance of hole carriers in the low temperature regime. This is further confirmed by the Hall data which we present later.   

    	\begin{table}	
    		\caption{Parameters obtained from a Dingle fitting of the SdH data shown in inset of Fig.~\ref{Fig-Hall} (a).}	
    				
    		\scalebox{1.2}{
    		\begin{tabular}{|c|c|c|c|} 
    			\hline
    			$T_D (K)$ & $\tau (10^{-13} s)$ &  l (nm) & $\mu (cm^2/V-s)$\\ [0.ex] 
    			\hline
    			2.62 & $4.6$ &  133& 2310\\ 
    			\hline
    		\end{tabular}}
    		\label{Table 2}
    	\end{table}

The amplitude of the quantum oscillations in resistivity and magnetization can be described by the equation 
    
    \begin{equation}
    \Delta \rho \propto - R_{T}R_{D} {\rm cos}(2\pi[{F\over B}-({1\over 2} + \beta + \delta)]) 
    \end{equation}  
    
    \begin{equation}
     \Delta M \propto - R_{T}R_{D} {\rm sin}(2\pi[{F\over B}-({1\over 2} + \beta + \delta)]) 
     \end{equation}
   	
\begin{figure*}[t]
	\centering
	\includegraphics[width=0.9\textwidth]{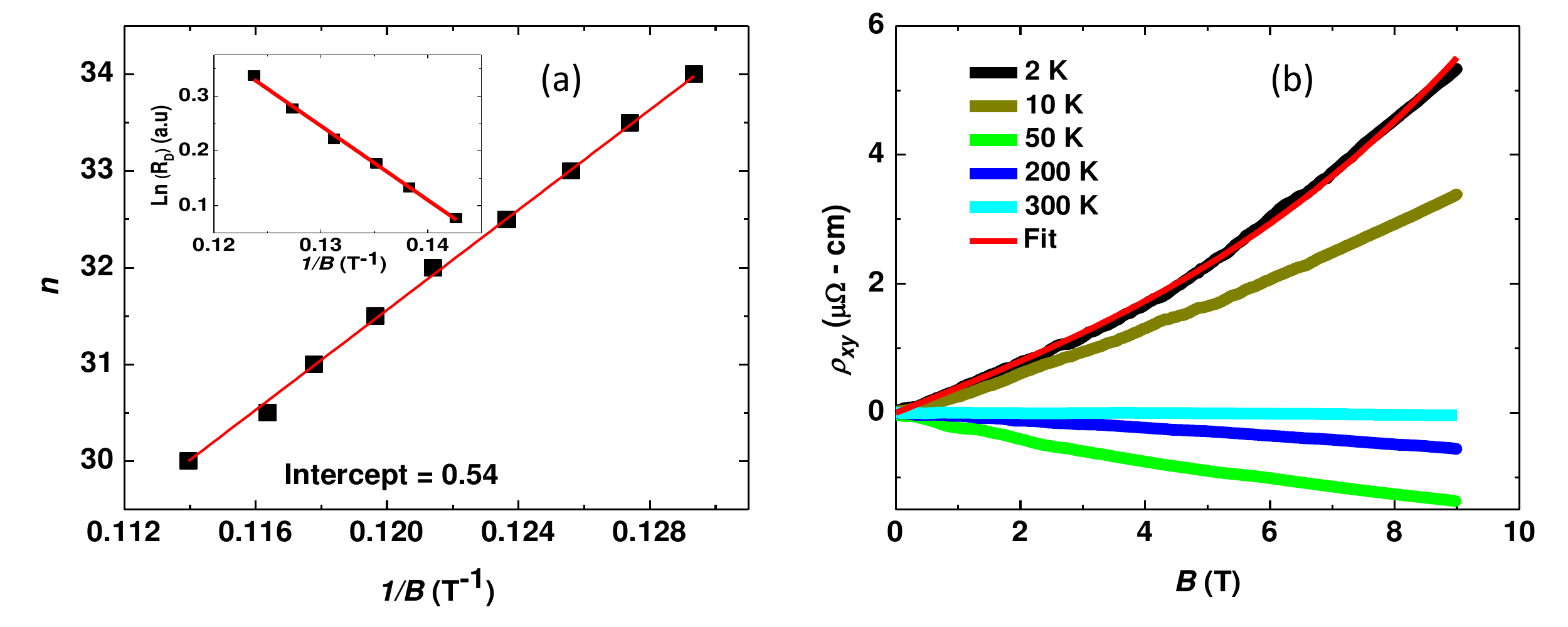}
	\caption{(a) Landau fan diagram extracted from the quantum oscillation data showing the Landau band index $n$ vs reciprocal of the field $1/B$.  The extrapolated intercept is close to what is expected ($0.5$) for a topologically non-trivial system.  (b) The Hall resistivity vs field at various temperatures.  The solid curve through the data at $T = 2$~K is a fit by a two-band model (se text for details).} 
	\label{Fig-Hall}
\end{figure*}

\noindent
where  $R_{T} = \lambda T/{\rm sinh}(\lambda T)$ and  $R_{D} = exp(-\lambda T_{D})$ are temperature and field induced damping factors respectively. Here, $\lambda = (2\pi^2 k_{B} m^*/\hbar eB)$, $m^*$ is the effective cyclotron mass, $T_{D}$ is the Dingle temperature and $2\pi\beta$ is the berry phase. The additional phase shift $\delta$ takes values of $0$ and $\pm 1/8$ for 2D and 3D systems, respectively\cite{34}. Fitting of the SdH oscillations with the above expressions gave an effective mass of $m^*$ = 0.35$m_0$ for the $\alpha$ pocket, where $m_0$ is the rest mass of electron.  A similar value $m^*$ = 0.32$m_0$, for the $\alpha$ pocket is obtained by fitting the dHvA oscillation amplitudes. In addition, the effective mass for the $\beta$ pocket is found to be 0.31$m_0$, which is comparable to the effective mass corresponding to the $\alpha$ pocket. These values of the effective mass for the two pockets are much higher (massive) than those observed in typical Dirac semimetals \cite{24,35} and are comparable to values found for other mono-pnictides in this series \cite{7, 11}.  Our results suggest that it is unlikely that the conduction in PrBi is primarily from the Dirac bands.  This would be consistent with ARPES studies which find the Dirac node to be located much below ($\approx 0.35$~eV) the Fermi level \cite{32}.
	
By fitting field dependent damping of the oscillation amplitude ($\alpha$ pocket)  with  $R_{D} = exp(-\lambda T_{D})$, we calculate a Dingle temperature of $T_{D}) = 2.62$~K\@.  The corresponding quantum life time is calculated to be $\tau = 4.6\times10^{-13}$~s and the quantum mobility is estimated to be $\mu = 2310~cm^2/$Vs. This value of the carrier mobility is higher than the value reported for LaBi \cite{7,36}.  A high value of carrier mobility has been suggested to be a signature of the Dirac carriers in LaBi and related compounds, but our data on PrBi shows that a high mobility can also be possible from three dimensional bands. Other parameters obtained from a Dingle fitting of the quantum oscillation data are given in Table~\ref{Table 2}. 
	   
In order to probe the possible topological nature of the bands in PrBi, we have constructed the Landau fan diagram from the quantum oscillation data as shown in Fig.~\ref{Fig-Hall}~(a). To construct the Landau fan diagram, we assign an integer value to the minima in the oscillation data of $\sigma_{xx}$. The extrapolated value of the intercept $n = 0.54(2)$ is close to the theoretical value ($0.5$) expected for a topologically non-trivial system.  However, it is worth mentioning that there will most likely be a large uncertainty in determining the intercept on the $n$ axis due to the large value of the Landau index ($30$) that we have to extrapolate from.    
	
Figure~\ref{Fig-Hall}~(b) shows the Hall resistivity at various temperatures.  At higher temperatures, the slope of the Hall resistivity is negative and almost linear, indicating that the dominant carriers are electrons at high temperature.  As the temperature is decreased below $10$~K, the slope of the Hall resistivity changes to a positive value with a weak nonlinearity at the higher fields.  The non-linearity becomes more pronounced at $T = 2$~K as shown in Fig.~\ref{Fig-Hall}~(b).  The change in the sign of the Hall resistivity indicates that there is a change in the dominant carrier type from electron to holes with decreasing temperature.  The nonlinear nature of the Hall resistivity also indicates that PrBi has contributions from two kinds of carriers.  For a quantitative analysis, the Hall resistivity data at $T = 2$~K was fitted by using a two band model \cite{31},
	
	\begin{equation}
	 \rho_{yx} = {B \over e} {(n_h\mu_h^2 - n_e\mu_e^2) + (n_h - n_e)(\mu_h\mu_e)^2B^2 \over (n_h\mu_h + n_e\mu_e)^2 + (n_h - n_e)^2(\mu_h\mu_e)^2B^2}~~~.
	\end{equation}

Here $n_h$, $\mu_h$ and $n_e$, $\mu_e$ are carrier density and mobility of holes and electrons respectively. The fit, shown as the solid curve through the $T = 2$~K data in Fig.~\ref{Fig-Hall} indicates that the electron and hole carrier density are approximately the same $\approx 0.45 \times 10^{19}~cm^{-3}$, indicating the nearly compensated semimetal nature of PrBi. The obtained mobilities of charge carriers are $3 \times 10^3 cm^2/$V-s.

\section{Theoretical Results}
In Table~\ref{DFT-a}, we compare the calculated lattice parameters of PrBi with experiments. We performed structural optimization (atoms positions and cell parameters were allowed to vary) to calculate cell parameters for the non-SOI case. For the calculations with SOI included, the lattice parameter $a$ was obtained by performing energy calculations at various constant $a$ values and these data were fit by the Birch-Murnaghan equation of state.  As can be seen from Table~\ref{DFT-a}, the GGA gives the unit-cell volume and lattice parameter to within about $1$\% of the experimental parameters for both non-SOI and SOI calculations. 

\begin{table}[h]
\caption{Calculated lattice parameters and unit cell volume compared with experimental values.}
\label{DFT-a}
\begin{tabular}{|c|c|c|c|}
\hline
Cell parameters&Non-SOI&SOI & expt \\  \hline
$a = b = c$ (\AA) &6.53 &6.54 &6.47\\ \hline
volume (\AA$^3$) &278.45 &279.73 &270.84\\ \hline
\end{tabular} 
 \end{table}

\subsection{Electronic band structure}
\begin{figure*}[t]
\includegraphics[width=\textwidth]{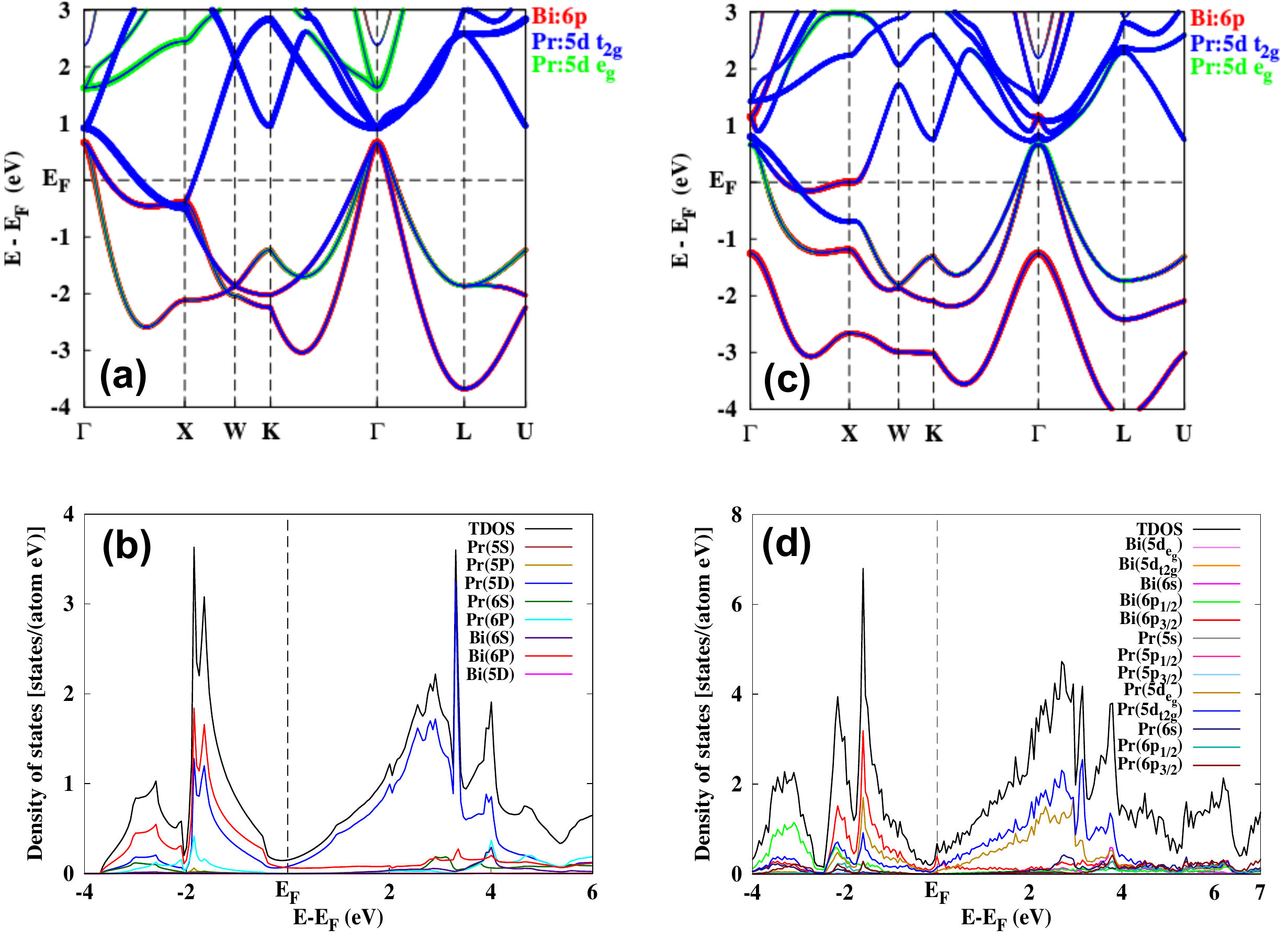}
\caption{(a) Electronic band structure and (b) Partial density of states (PDOS) for non-spin-orbit calculation.  (c) Electronic band structure and (d) Partial density of states (PDOS) with spin-orbit included.}
\label{DFT1} 
\end{figure*}

The electronic structure and partial density of states of PrBi without and with spin-orbit coupling are shown in Fig.~\ref{DFT1}~(a), (b) and (c), (d), respectively. Figure~\ref{DFT1} illustrates the semi-metallic behavior of PrBi with a small density of states at the Fermi level. For both non spin-orbit and spin-orbit calculations we find that, close to the $\Gamma$ and X high-symmetry points of the Brillouin zone, energy bands cross the Fermi level (E$_F$), creating hole pockets at the $\Gamma$ point and electron pockets at the X points. For the non-spin-orbit case, four bands cross E$_F$, with three hole-like bands at $\Gamma$ point and one electron-like band at X point as can be seen in Fig.~\ref{DFT1}(a).  The inclusion of spin-orbit coupling splits the bands and now there are three doubly-degenerate bands crossing the E$_F$.  Out of these three bands, two are hole-like near $\Gamma$ and one is electron-like near X as can be seen in Fig.~\ref{DFT1}(c). 

As can be seen from Fig.\ref{DFT1}(b), for non-SOI calculations, near the Fermi level the valence band is primarily a mixture of Bi $p$-orbitals.  On the other hand, the conduction band mainly consists of Pr $d$-orbitals with small contribution of Bi $p$-orbitals.  The strong spin-orbit interaction splits degenerate Bi $p$-orbitals into $p_{3/2}$ and $p_{1/2}$ orbitals with a splitting of $\approx 1.9$~eV.  The $p_{1/2}$ orbital goes into the valence band around $-1.2$~eV and the $p_{3/2}$ orbitals contribute near the Fermi level and the valence band. The Pr $d$-orbitals split into $e_g$ and $t_{2g}$, and SOI further splits $t_{2g}$ into doubly degenerate and non-degenerate orbitals. 

The band-structure between $\Gamma$ and X has some important features.  For the non SOI case shown in Fig.~\ref{DFT1}(a) Bi($p$) and Pr($d$) orbitals cross close to the X point between $\Gamma$ and X\@.  With SOI, there is a band inversion below E$_F$ along $\Gamma$-X creating a hole pocket and a gap of $0.7$~eV opens up at the X point.  The direct band gap at $\Gamma$ point reduces from $0.26$~eV for non-SOI to $0.15$~eV for SOI. The overlap between the valence band at $\Gamma$ and conduction band at X is $0.65$~eV for PrBi with SOI. 
The electronic band structure of PrBi with and without spin-orbit interactions are consistent with a recent report using VASP implementation of DFT \cite{Chen2018}. 

\begin{figure*}[t]
\includegraphics[width=0.8\textwidth]{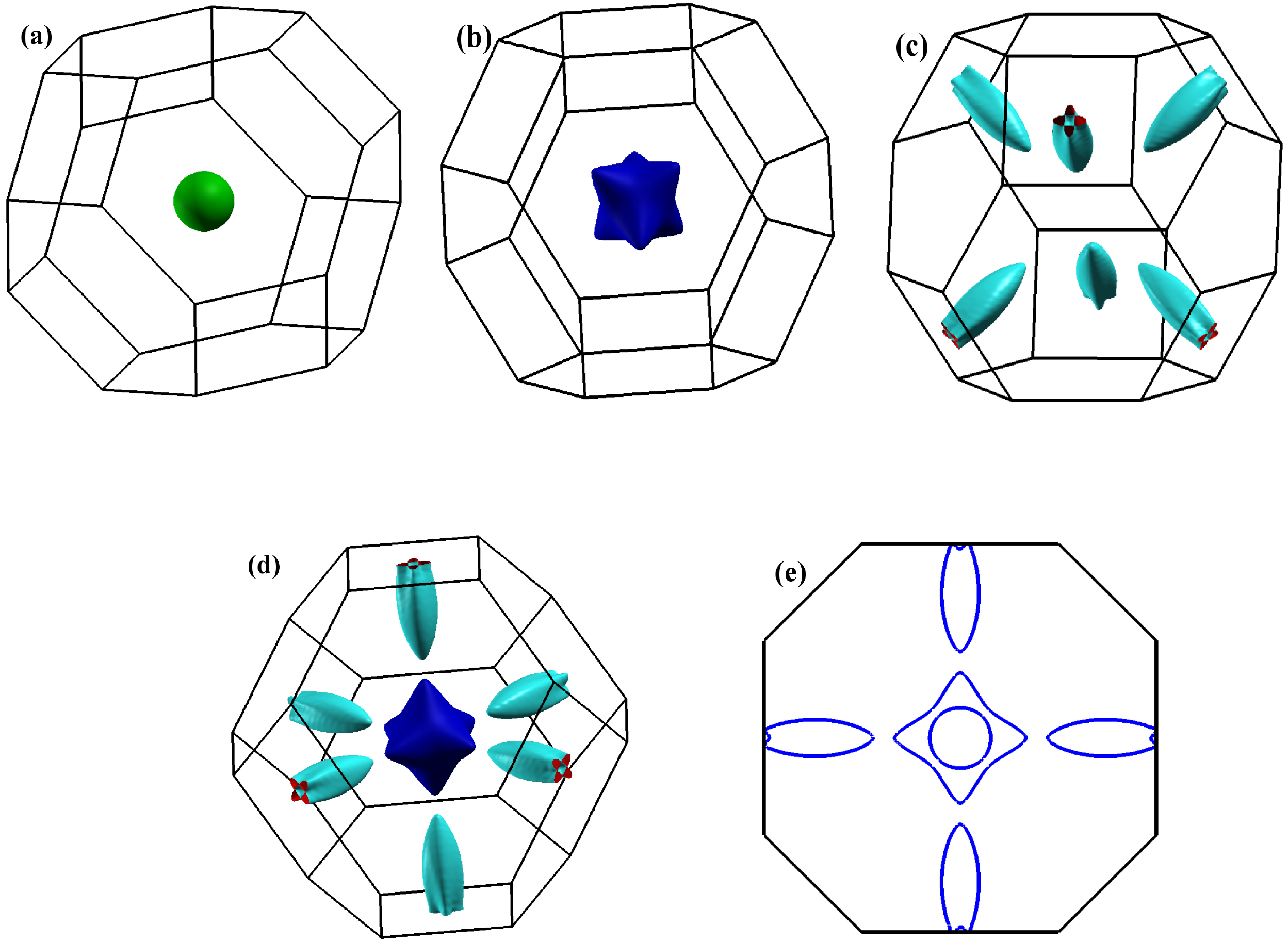}
 \caption{(a)-(c) The Fermi surface of PrBi in the first Brillouin zone for each band crossing the Fermi energy. (d) The total Fermi surface including all crossing bands. (e) The Fermi surface slice for (001) plane. In (a) and (b) we show the two hole pockets of the FS at the $\Gamma$ point and in (c) we show the electron pockets at each $X$ point of the BZ.  }
 \label{fig-FS}
 \end{figure*}

\subsection{Fermi surface}
Figure~\ref{fig-FS} displays the Fermi surface (FS) of PrBi with SOI. DFT predicts a spherically symmetric and a star-shaped hole-pocket at $\Gamma$ shown in Fig.~\ref{fig-FS}(a) and (b), respectively. Additionally, an electron-pocket is found at X as shown in Fig ~\ref{fig-FS}(c).  The three FS pockets are consistent with three quantum oscillation frequencies observed in experiments.  The combined FS and a slice of the FS along the (001) plane are shown in Fig.~ \label{fig-FS}(d) and Fig.~ \label{fig-FS}(e), respectively. To compare with experiments, in Table~\ref{DFT-b} we list the dHvA frequencies predicted by DFT for both with and without inclusion of spin-orbit interaction on DFT-optimized and experimental lattice parameters. The magnetic field $H$ is parallel to the $z$-axis. To calculate dHvA frequencies we have used the maximally localized Wannier function to generate the Fermi surface. From Table~\ref{DFT-b} we note that the dHvA frequencies have a significant dependence on spin-orbit interaction and volume. So in order to compare with experiment it is very important the DFT calculations capture the correct picture of spin-orbit interaction.   We observe that the quantum oscillation frequencies $\alpha$ and $\beta$ are overestimated and $\gamma$ is underestimated in DFT calculations. However, the calculated quantum oscillation frequencies of
PrBi without spin-orbit interaction agree fairly well with the experimental data.

 \begin{table}[t]
 \caption{Comparison between DFT-calculated and experimentally observed de Haas-van Alphen (dHvA)/Shubnikov de Haas (SdH) frequencies in (kT).  The $\alpha$ is the electron band near X, and $\beta$ and $\gamma$ are the hole pockets near $\Gamma$.  DFT without SOI predicts an extra hole-pocket ($\beta_2$) near $\Gamma$.  Effective mass m$^*$ in m$_e$ is given in parenthesis.}
 \begin{tabular}{|c|cccc|}
 \hline
 &&DFT-Calculated & &Experiment\\ \hline
 &with-SOI & no-SOI& no-SOI&\\    
 & Optimized & Exp-lattice & Optimized &  \\ \hline
$\alpha$ &0.42(0.47)&0.58(0.57)&0.30 (0.15)& 0.264\\
$\beta$ &0.67(0.16) & 0.68(0.16)&0.52(0.10)& 0.57\\
$\gamma$ & 1.3 (0.34)&1.27(0.32)&0.11(0.23)& 1.52\\
$\beta_2$ &-&-&1.47 (0.34)&\\ \hline
\end{tabular} 
\label{DFT-b}
\end{table}

\section{Conclusion}
We have presented a detailed magneto-transport study on high quality single crystals of the mono-pnictide PrBi.  Magnetic measurements are consistent with a localized trivalent Pr$^{3+}$ valence state and reveal paramagnetic temperature dependent behaviour down to a temperature of $T = 2$~K\@.  Thus PrBi is a potential material to observe Topological properties in the presence of electronic correlations.  Hall measurements reveal a two-carrier compensated nature of transport in PrBi.  The temperature dependent resistivity in zero magnetic field shows a typical semimetallic character and a field induced metal insulator transition is observed at low temperatures $T \leq 20$~K\@.  We observe a large non-saturating magnetoresistance ($\approx 4.4 \times 10^4~\%$) at low temperatures ($T= 2$~K) in a magnetic field $B = 9$~T\@.  Additionally, pronounced quantum oscillations are observed in the magnetoresistance and magnetization data.  We map out the Fermi surface (FS) of PrBi by angle dependent SdH and dHvA quantum oscillation measurements which reveal a three dimensional FS made up of two hole bands at the $\Gamma$ point and an electron band at the X point in the BZ.  Our magnetotransport measurements reveal effective masses for the $\alpha$ and $\beta$ Fermi surface pockets which are quite large and unlikely to arise from Dirac bands.  The high mobility that we estimate is therefore somewhat surprising since a high mobility has usually been connected to transport of Dirac electrons in Topological materials.  The three FS pockets observed in quantum oscillations are consistent with our electronic-structure calculations.  These results are consistent with recent ARPES measurements \cite{22}.  The electronic-structure calculations also reveal a band inversion along the $\Gamma$-X directions leading to expectations of Topologically non-trivial effects.  A Landau fan diagram analysis gives a value of $\pi$ for the Berry phase which suggests Topologically non-trivial charge carriers.  However, in the Landau fan diagram analysis we had to extrapolate down from a large Landau level index $n$ in our case which means that the intercept will have a large error.  Future quantum oscillation measurements to higher fields will be required to get a more reliable estimate of the Berry phase.

\emph{Acknowledgments.--} DS acknowledges the Finnish IT Center for Science and Inter-University Accelerator Center, New Delhi, India for providing computing resources.

\noindent
$^*$ These authors contributed equally to this work.


\begin{references}
\bibitem{1} M. Z. Hasan and C. L. Kane, Rev. Mod. Phys. {\bf 82}, 3045 (2010).

\bibitem{2} B. A. Bernevig, T. L. Hughes, and S.-C. Zhang, Science {\bf 314}, 1757 (2006),

\bibitem{3} X. Wan, A. M. Turner, A. Vishwanath, and S. Y. Savrasov, Phys. Rev. B {\bf 83}, 205101 (2011).

\bibitem{4} A. A. Burkov and L. Balents, Phys. Rev. Lett. {\bf 107}, 127205 (2011).

\bibitem{5} S. M. Young, S. Zaheer, J. C. Y. Teo, C. L. Kane, E. J. Mele, and A. M. Rappe, Phys. Rev. Lett. {\bf 108}, 140405 (2012).

\bibitem{6} Z. K. Liu, B. Zhou, Y. Zhang, Z. J. Wang, H. M. Weng, D. Prabhakaran, S.-K. Mo, Z. X. Shen, Z. Fang, X. Dai, Z. Hussain, and Y. L. Chen, Science {\bf 343}, 864 (2014).

\bibitem{20} G. Li, Z. Xiang, F. Yu, T. Asaba, B. Lawson, P. Cai, C. Tinsman, A. Berkley, S. Wolgast, Y. S. Eo, D.-J. Kim, C. Kurdak, J. W. Allen, K. Sun, X. H. Chen, Y. Y. Wang, Z. Fisk, and L. Li, Science {\bf 346}, 1208 (2014),

\bibitem{Wan2011} X. Wan, A. M. Turner, A. Vishwanath, and Sergey Y. Savrasov, Phys. Rev. B {\bf 83}, 205101 (2011).

\bibitem{7} F. F. Tafti, Q. D. Gibson, S. K. Kushwaha, N. Haldolaarachchige, and R. J. Cava, Nature Physics {\bf 12}, 272 (2015).

\bibitem{8} L.-K. Zeng, R. Lou, D.-S. Wu, Q. N. Xu, P.-J. Guo, L.-Y. Kong, Y.-G. Zhong, J.-Z. Ma, B.-B. Fu, P. Richard, P. Wang, G. T. Liu, L. Lu, Y.-B. Huang, C. Fang, S.-S. Sun, Q. Wang, L. Wang, Y.-G. Shi, H. M. Weng, H.-C. Lei, K. Liu, S.-C. Wang, T. Qian, J.-L. Luo, and H. Ding, Phys. Rev. Lett. {\bf 117}, 127204 (2016).

\bibitem{9} N. Kumar, C. Shekhar, S.-C. Wu, I. Leermakers, O. Young, U. Zeitler, B. Yan, and C. Felser, Phys. Rev. B {\bf 93}, 241106(R) (2016).

\bibitem{10} N. Wakeham, E. D. Bauer, M. Neupane, and F. Ronning, Phys. Rev. B {\bf 93}, 205152 (2016).

\bibitem{11} F. Wu, C. Y. Guo, M. Smidman, J. L. Zhang, and H. Q. Yuan, Phys. Rev. B {\bf 96}, 125122 (2017).

\bibitem{12} S. Sun, Q. Wang, P.-J. Guo, K. Liu, and H. Lei, New Journal of Physics {\bf 18}, 082002 (2016).

\bibitem{13} J. Nayak, S.-C. Wu, N. Kumar, C. Shekhar, S. Singh, J. Fink, E. E. D. Rienks, G. H. Fecher, S. S. P. Parkin, B. Yan, and C. Felser, Nature Communications {\bf 8}, 13942 (2017).

\bibitem{14} B. Feng, J. Cao, M. Yang, Y. Feng, S. Wu, B. Fu, M. Arita, K. Miyamoto, S. He, K. Shimada, Y. Shi, T. Okuda, and Y. Yao, Phys. Rev. B {\bf 97}, 155153 (2018).

\bibitem{15} H.-Y. Yang, T. Nummy, H. Li, S. Jaszewski, M. Abramchuk, D. S. Dessau, and F. Tafti, Phys. Rev. B {\bf 96}, 235128 (2017).

\bibitem{16} Y. J. Hu, E. I. Paredes Aulestia, K. F. Tse, C. N. Kuo, J. Y. Zhu, C. S. Lue, K. T. Lai, and S. K. Goh, Phys.Rev.B {\bf 98}, 035133 (2018).

\bibitem{17} J. Xu, N. J. Ghimire, J. S. Jiang, Z. L. Xiao, A. S. Botana, Y. L. Wang, Y. Hao, J. E. Pearson, and W. K. Kwok, Phys. Rev. B {\bf 96}, 075159 (2017).

\bibitem{18} Y.-Y. Wang, H. Zhang, X.-Q. Lu, L.-L. Sun, S. Xu, Z.-Y. Lu, K. Liu, S. Zhou, and T.-L. Xia, Phys. Rev. B {\bf 97}, 085137 (2018).

\bibitem{19} O. Pavlosiuk, P. Swatek, D. Kaczorowski, and P. Wisniewski, Phys. Rev. B {\bf 97}, 235132 (2018).

\bibitem{Tsuchida-a} T. Tsuchida and W. E. Wallace, J. Chem. Phys. {\bf 43}, 2087 (1965).

\bibitem{Tsuchida-b} T. Tsuchida and W. E. Wallace, J. Chem. Phys. {\bf 43}, 2885 (1965).

\bibitem{Giannozzi2009} Paolo Giannozzi et al. J. of Phys.: Conden. Matt., {\bf 21} 395502 (2009).

\bibitem{Giannozzi2017} P Giannozzi et al., Journal of Physics: Condensed Matter, 29(46):465901, 2017.

\bibitem{PBE}  J. P. Perdew, K. Burke, and M. Ernzerhof, Phys. Rev. Lett., {\bf 77}, 3865 (1996).

\bibitem{Kokalj2003} A. Kokalj, Comp. Mater. Sci., {\bf 28}, 155 (2003).

\bibitem{Pizzi2014} G Pizzi, YS Lee I Souza D Vanderbilt N Marzari A A Mostofi, JR Yates, Comput. Phys. Commun., {\bf 185}, 2309 (2014).

\bibitem{Rourke2012} P. M. C. Rourke and S. R. Julian, Comput. Phys. Commun., {\bf 183}, 324 (2012).

\bibitem{Chen2018} Jia Chen, Xu Duan, Fan Wu. Communication Physics, 1(71):1?7, 2018.

\bibitem{21} X. H. Niu, D. F. Xu, Y. H. Bai, Q. Song, X. P. Shen, B. P. Xie, Z. Sun, Y. B. Huang, D. C. Peets, and D. L. Feng, Phys. Rev. B {\bf 94}, 165163 (2016).

\bibitem{22} P. Li, Z. Wu, F. Wu, C. Cao, C. Guo, Y. Wu, Y. Liu, Z. Sun, C.-M. Cheng, D.-S. Lin, F. Steglich, H. Yuan, T.-C. Chiang, and Y. Liu, Phys. Rev. B {\bf 98}, 085103 (2018).

\bibitem{23} M. N. Ali, J. Xiong, S. Flynn, J. Tao, Q. D. Gibson, L. M. Schoop, T. Liang, N. Haldolaarachchige, M. Hirschberger, N. P. Ong, and R. J. Cava, Nature {\bf 514}, 205 EP (2014).

\bibitem{24} C. Shekhar, A. K. Nayak, Y. Sun, M. Schmidt, M. Nicklas, I. Leermakers, U. Zeitler, Y. Skourski, J. Wosnitza, Z. Liu, Y. Chen, W. Schnelle, H. Borrmann, Y. Grin, C. Felser, and B. Yan, Nature Physics {\bf 11}, 645 (2015).

\bibitem{25} C.-L. Zhang, Z. Yuan, Q.-D. Jiang, B. Tong, C. Zhang, X. C. Xie, and S. Jia, Phys. Rev. B {\bf 95}, 085202 (2017).

\bibitem{26} Y.-Y. Wang, H. Zhang, X.-Q. Lu, L.-L. Sun, S. Xu, Z.-Y. Lu, K. Liu, S. Zhou, and T.-L. Xia, Phys. Rev. B {\bf 97}, 085137 (2018).

\bibitem{27} S. Nimori, G. Kido, D. Li, and T. Suzuki, Physica B: Condensed Matter {\bf 211}, 148 (1995), 

\bibitem{28} L. Ye, T. Suzuki, C. R. Wicker, and J. G. Checkelsky, Phys. Rev. B {\bf 97}, 081108(R) (2018).

\bibitem{30} Y. Ando, Journal of the Physical Society of Japan {\bf 82}, 102001 (2013).
\bibitem{31} C. M. Hurd, The Hall Effect in Metals and Alloys (Plenum Press, 1972), New York.
\bibitem{32} P. Monachesi, Z. Domanski, and M. S. S. Brooks, Phys. Rev. B {\bf 50}, 1013 (1994).
\bibitem{34} C. M. Wang, H.-Z. Lu, and S.-Q. Shen, Phys. Rev. Lett. {\bf 117}, 077201 (2016).
\bibitem{35} T. Liang, Q. Gibson, M. N. Ali, M. Liu, R. Cava, and N. Ong, Nature materials {\bf 14}, 280 (2015).
\bibitem{36} R. Singha, B. Satpati, and P. Mandal, Scientific Reports {\bf 7}, 6321 (2017).

\bibitem{Chen2018} J. Chen, Xu Duan, Fan Wu. Communication Physics, {\bf 1}, 71 (2018).
\end{references}
\end{document}